\documentclass[aps,twocolumn,showpacs,preprintnumbers,amsmath,amssymb,%
floatfix,prl,superscriptaddress]{revtex4}
\usepackage[dvips]{graphicx}

\newcommand{\mured}{\mu_{\text{red}}}
\newcommand{\Zeff}{Z_{\text{eff}}}
\newcommand{\Reff}{R_{\text{eff}}}
\newcommand{\tildeZ}{\tilde{Z}}
\newcommand{\tildeZeff}{\tilde{Z}_{\text{eff}}}

\begin{document}

\title{Electrophoresis of colloidal dispersions in the
low-salt regime}

\author{Vladimir Lobaskin}
\affiliation{Max Planck Institute for Polymer Research,
Ackermannweg 10, D-55128 Mainz, Germany}
\affiliation{Physik-Department, Technische Universit\"at Munchen,
James--Franck--Stra{\ss}e, D-85747 Garching, Germany}

\author{Burkhard D\"unweg}
\affiliation{Max Planck Institute for Polymer Research, Ackermannweg
10, D-55128 Mainz, Germany}

\author{Martin Medebach}
\affiliation{Institut f\"ur Physik, Johannes
Gutenberg--Universit\"at, Staudingerweg 7, D-55128 Mainz, Germany}

\author{Thomas Palberg}
\affiliation{Institut f\"ur Physik, Johannes
Gutenberg--Universit\"at, Staudingerweg 7, D-55128 Mainz, Germany}

\author{Christian Holm}
\affiliation{Max Planck Institute for Polymer Research, Ackermannweg
10, D-55128 Mainz, Germany}
\affiliation{Frankfurt Institute for Advanced Studies (FIAS),
J.~W. Goethe--Universit\"at, Max--von--Laue--Stra{\ss}e 1,
D-60438 Frankfurt/Main, Germany}

\date{Revised \today}
\begin{abstract}
  We study the electrophoretic mobility of spherical charged colloids
  in a low-salt suspension as a function of the colloidal
  concentration. Using an effective particle charge and a reduced
  screening parameter, we map the data for systems with different
  particle charges and sizes, including numerical simulation data with
  full electrostatics and hydrodynamics and experimental data for
  latex dispersions, on a single master curve. We observe two
  different volume fraction-dependent regimes for the electrophoretic
  mobility that can be explained in terms of the static properties of
  the ionic double layer.
\end{abstract}


\maketitle

Many important properties of colloidal dispersions are directly or
indirectly determined by the electric charge of the colloidal
particles. Phase stability is provided by the repulsive
interaction between like charges, while the details of the static
and dynamic behavior are the result of the interplay between
electrostatic interactions between macroions, counterions, and
added salt ions, the dielectric response of the solvent, and the
solvent hydrodynamics. Depending on whether the concentration of
the background electrolyte (relative to that of the ``native''
counterions) is large or small, one must expect quite different
behavior both with respect to statics and dynamics. The
salt-dominated regime (high salt concentration) has been studied
extensively both for static \cite{HansenRev} and dynamic
properties \cite{wiersema66a,obrien78a,lozada01a,Carrique03}. In
this case, the large reservoir of salt ions results in a strong
Debye screening of the electrostatic interactions between the
macroions, such that the net charge density is essentially zero
throughout the dispersion except for narrow ionic atmospheres
around the colloids. An external electric field will therefore
exert forces only within these layers, such that hydrodynamic
interactions are also strongly screened in a system subjected to
electrophoresis \cite{Long,tanaka01a}. Therefore, the problems
both of ion cloud structure and of electrophoresis can be treated
within a single-macroion framework \cite{Russel}, and the
dependence of the electrophoretic mobility $\mu = v / E$ ($v$
denoting the colloid drift velocity and $E$ the driving electric
field) on the macroion volume fraction $\Phi$ is weak.

Conversely, in the interesting regime of low salt concentration,
where this screening is not present, the concentration dependence
can be significant \cite{Ohshima,Yamamoto}. From the theoretical
point of view, this regime remains practically unexplored. The
main goal of the present Letter is to find general relations
between the electrophoretic mobility and other dispersion
parameters in the counterion-dominated regime, based upon data
from both computer simulations and experiments.

An important reference point in the case of low salt is the H\"uckel
limit of electrokinetics. For an isolated colloidal sphere ($\Phi =
0$, salt concentration zero) with radius $R$ and charge $Z e$ ($e$
denoting the elementary charge) in a solvent of viscosity $\eta$ and
dielectric constant $\epsilon$, Stokes' law implies $\mu = Z e / (6
\pi \eta R)$ (note that all counterions are infinitely far away).
Furthermore, we introduce the Bjerrum length $l_B = e^2 / (4 \pi
\epsilon k_B T)$, $k_B$ and $T$ denoting Boltzmann's constant and the
temperature, respectively, as a characteristic length scale of
electrostatic interactions. The reduced (dimensionless) mobility,
defined as $\mured = 6 \pi \eta l_B \mu / e$, thus assumes the
value $\mured = Z l_B / R$ in the H\"uckel limit.

We are now interested in the case $\Phi > 0$ (but still
salt-free). Some non-trivial statements about this regime can be
made already in terms of dimensional analysis. $\mured$, as a
dimensionless quantity, can only depend on dimensionless
combinations of the essential parameters of the system. These are:
(i) $k_B T$ as a typical energy scale; (ii) $l_B$ as the
fundamental length scale of electrostatics ($l_B$ may be viewed as
re-parametrization of $\epsilon$); (iii) the solvent viscosity
$\eta$ (or alternatively $\mu_0 = e / (6 \pi \eta l_B)$)
providing the fundamental time scale for viscous dissipation; (iv)
the colloid radius $R$ and (v) its valence $Z$; (vi) the colloid
volume fraction $\Phi$; and (vii) the radius $a$ of the
(monovalent) counterions (which we consider, due to Stokes' law,
also as a measure of their mobility). Using the first three
parameters as those which provide our fundamental unit system for
energy, length, and time, we thus find $\mured = \mured (\Phi, Z,
l_B / R, l_B / a)$.

We now introduce two further re-parametrizations. Firstly, we
replace $Z$ by $\tildeZ = Z l_B / R$, which would be equal to
$\mured$ in the $\Phi = 0$ limit. We expect that this change of
variables results in a much weaker dependence on the last two
parameters. This is obvious for $\Phi = 0$ where there is no
dependence left; note also that $l_B / a$ is typically of order
unity for most systems. Secondly, in order to make contact with
the standard parametrization used in the high-salt case
\cite{Russel,Hunter}, we use the variable $\kappa R$ instead of
$\Phi$, where $\kappa$ has the dimension of an inverse length, and
$\left( \kappa R \right)^2 = 3 \tildeZ \Phi$. Since the average
counterion concentration is $n_{i} = Z / V$, where $V$ is the
average system volume per colloidal sphere, and $\Phi = 4 \pi R^3
/(3 V)$, one sees that $\kappa^2 = 4 \pi l_B n_{i}$, i.~e.
$\kappa$ would have the interpretation of a Debye screening
parameter if $n_{i}$ were the concentration of salt ions. We would
like to stress, however, that these are mere re-parametrizations.
This implies neither the interpretation of $\tildeZ$ as an
electrostatic potential, nor of $\kappa$ in terms of Debye
screening. With these caveats in mind, we will call $\tildeZ$ the
reduced charge, and $\kappa$ the screening parameter
\cite{Belloni}. From these considerations we expect that
measurements for different systems should yield equal $\mured$
values if the physical situations are identical in terms of the
two parameters $\kappa R$ (or $\Phi$), and $\tildeZ$. As we will
show below, this is indeed the case.

As a final crucial step, we introduce \emph{effective} values for
the parameters $Z$ and $R$ (and correspondingly also $\Phi$) and
the effective reduced charge $\tildeZeff = \Zeff l_B / \Reff$.
While counterions far away from the colloid can be described
reasonably well in terms of small Stokes spheres, co-moving with
the surrounding hydrodynamic flow, this is much less obvious for
those ions in the close vicinity of the colloid, which are tightly
coupled to its motion. This is particularly true for highly
charged systems. For this reason, we combine the central colloid
with some of its counterions to an effective sphere with slightly
increased $R$ as well as decreased $Z$. $\Reff$, the effective
radius in the computer model, was defined as the minimum distance
between the center of the colloid and the center of a counterion
(see below). In the experiment, we set $\Reff = R$ due to the
smallness of the counterions. For determining $\Zeff$ both in
simulation and experiment, we applied the concept of charge
renormalization \cite{chargerenormalization,Belloni,Trizac}: After
obtaining the full numerical solution of the Poisson-Boltzmann
(PB) equation within the framework of the Wigner-Seitz cell model,
this function was fitted to the solution of the corresponding
linearized equation near the cell boundary \cite{Belloni}.
Correspondingly, the screening parameter has to be calculated now
with the effective counterion concentration, i.~e. $n_{i} =
\Zeff/V$ for the salt-free system. The comparison between
simulation and experiment is then done in terms of these effective
parameters. The degree of charge renormalization is weak for small
$Z$ but is quite substantial for larger values (see below).

\begin{figure}
\includegraphics[clip, width=8cm]{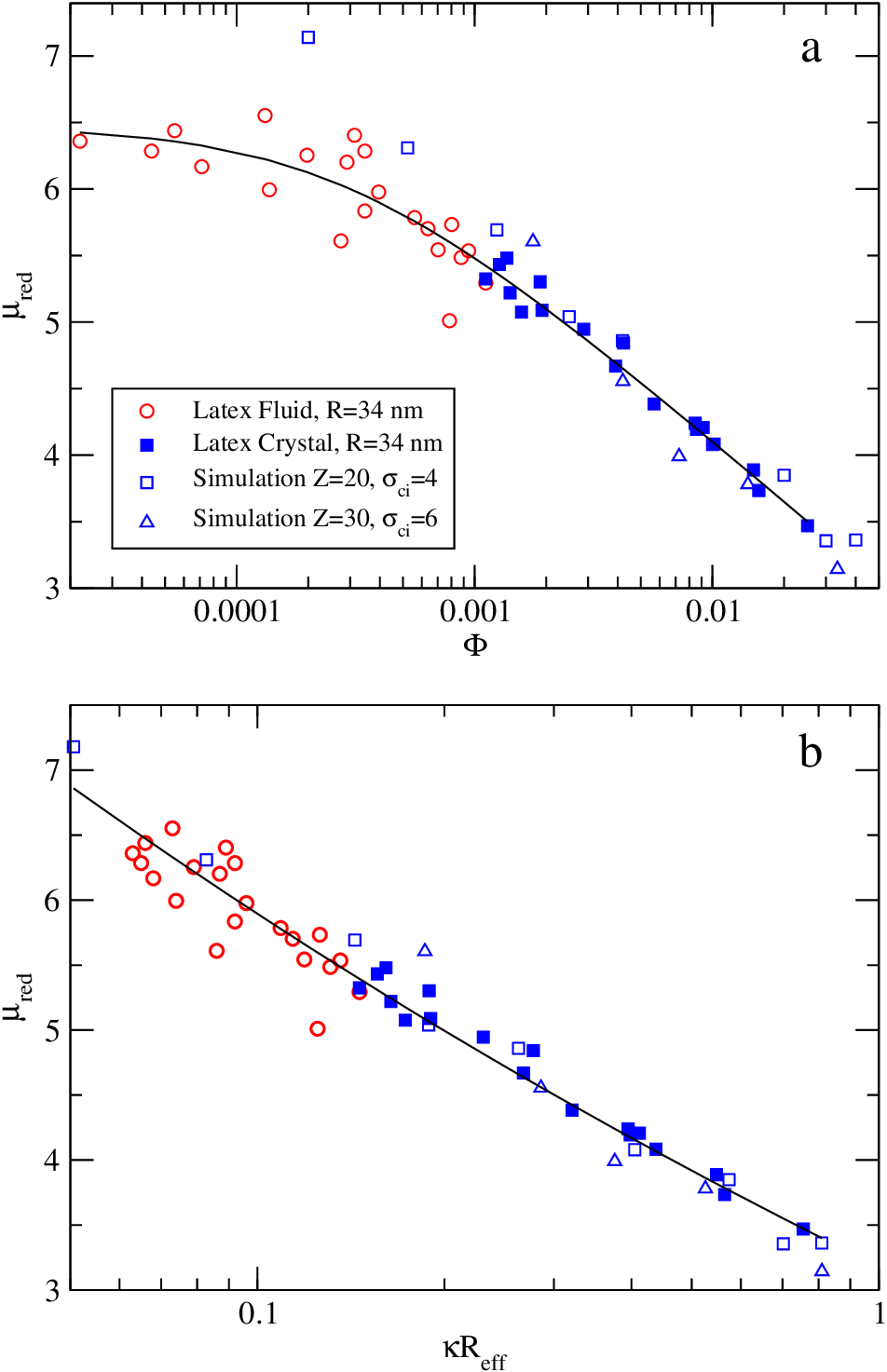}
\caption{Reduced electrophoretic mobility of a spherical
  particle vs. (a) particle volume fraction, and (b) reduced screening
  parameter, calculated from counterion and salt contributions (see text)
  for various systems with matching $\tildeZ$.
  The solid curves represent a fit to the experimental data.
  } \label{fig:salt1}
\end{figure}

Our Molecular Dynamics (MD) computer model comprises small
counterions, plus one large charged sphere, around which we wrap a
network of small particles. All small particles are coupled
dissipatively to a Lattice Boltzmann (LB) background
\cite{Ahlrichs1,Ahlrichs2}, which provides the hydrodynamic
interactions. Thermal motion is taken into account via Langevin
noise, and electrostatic interactions are calculated via the Ewald
summation technique. The system is simulated in a box with
periodic boundary conditions, and thus corresponds to a
well-defined finite value of $\Phi$. In our unit system, we use
$l_B = 1.3$, $R = 3$ or $5$ and $Z = 20$, $30$, or $60$, while the
small monovalent ions have diameter unity, which is also the LB
lattice spacing. The closest colloid-ion separation was set to
$\sigma_{ci} = R + 1$, which was also taken as the effective
macroion radius, $\Reff = \sigma_{ci}$. For further details, see
Refs.~\cite{lobaskin99,NJP,JPCM}. Applying an external electric
field, and averaging the steady-state velocity of the colloid, we
calculated $\mu$. Some data points were compared with zero-field
data obtained from Green-Kubo integration, confirming that the
response is still in the linear regime. Due to the ``screening''
of hydrodynamic interactions (the total force on the system
vanishes, due to charge neutrality, and thus does not generate a
large-scale flow) \cite{Long,tanaka01a}, we expect (and find, see
below) only weak finite-size effects.

The experimental system comprises thoroughly deionized aqueous
suspensions of latex spheres (lab code PnBAPS68, kindly provided
by BASF, Ludwigshafen, Germany) with diameter $2R = 68 \text{nm}$,
low polydispersity and high charge (bare charge $Z \approx 1500$;
effective charge from conductivity $Z^*_\sigma = 450$
\cite{tom2}). They show a low lying and narrow first order
freezing transition at a particle number density of $n_F \approx
(6.1 \pm 0.3) \mu \text{m}^{-3}$. Due to the small size optical
investigations are possible without multiple scattering up to
large $n$. Using Doppler velocimetry in the super-heterodyning
mode \cite{tom4} we studied the electrophoretic mobility in the
range of $0.1 \mu \text{m}^{-3} \leq n \leq 160 \mu
\text{m}^{-3}$, corresponding to volume fractions $\Phi =
(4\pi/3)R^3 n$ of $5 \times 10^{-4} \leq \Phi \leq 2.5 \times
10^{-2}$ \cite{tom5}. Consistent with data taken on other species
and also at elevated salt concentrations \cite{tom6}, the mobility
as shown in Fig. \ref{fig:salt1} exhibits a plateau at low $\Phi$
and descends fairly linearly in this semi-logarithmic plot at
larger $\Phi$. Charge renormalization \cite{Belloni} yields weakly
varying values for $\Zeff$, resulting in $\tildeZeff \approx 8.5$
\cite{tom7}. In the simulations we chose the parameter sets $(Z =
20, R = 3)$ and $(Z = 30, R = 5)$; these correspond to roughly the
same value of $\tildeZeff$ in the dilute limit $\Phi \to 0$.

Comparing the experimental and simulated mobility data in Fig.
\ref{fig:salt1}a, where $\mured$ is shown as a function of $\Phi$,
one observes good agreement as long as $\Phi$ is not too small.
However, simulation and experiment deviate in the regime of very
low volume fractions. The reason is that for very small $\Phi$ the
dissociation of water starts to play a role in the experiment ---
the size of the counterion cloud is no longer determined by the
colloid-colloid distance, but rather by the background ionic
concentration, so that it remains finite even at $\Phi=0$. In
principle, one must expect that the salt species will introduce
yet another dimensionless parameter into the problem. However, we
found that the effect of salt can be incorporated, within a
reasonable approximation, by just adding the salt concentration to
the counterion concentration, such that we obtain a new scaling
variable $\kappa \Reff$ with $\kappa^2 = 4 \pi l_B \left(n_{i} +
n_{\text{salt}} \right)$. The use of this procedure can be
supported by the observation that the electrostatic potential in
the regions centered between the colloids varies only weakly, such
that a description in terms of a linearized PB equation is
possible. In these regions, however, the local counterion
concentration contributes to the screening parameter, too. Figure
\ref{fig:salt1}b shows that this strategy to include the effect of
salt is indeed successful.

\begin{figure}
\includegraphics[clip, width=7.5cm]{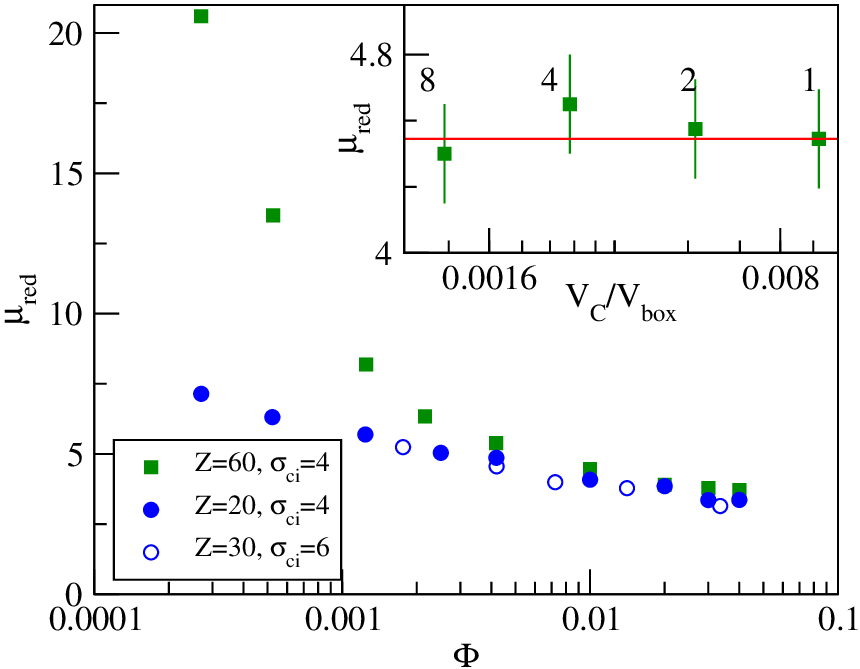}
\caption{Electrophoretic mobility of a spherical particle of the
 indicated charge and size as a function of the particle volume
 fraction in a salt--free system. The effective reduced charge
 $\tildeZeff$ is equal for the three systems at
 $\Phi \geq 0.01$, while it is growing in the system with $Z=60$
 upon stronger dilution. The inset shows the mobility of the particle
 of $Z = 60$ as a
 function of $V_C / V_{box}$, where $V_C$ is the volume of one
 colloidal particle, and $V_{box}$ is the system volume at constant
$\Phi$. } \label{fig:phi}
\end{figure}

A qualitatively different behavior of $\mured$ is illustrated in
Fig.~\ref{fig:phi}, which shows the same simulation data as those
of Fig.~\ref{fig:salt1}, augmented by an additional data set
obtained at $Z = 60$. Upon stronger dilution $\mured$ for the
system with $Z=60$ increases sharply as $\Phi \to 0$ due to
evaporation of the condensed ion layer, i.~e. increase of $\Zeff$.
In this regime, we find empirically $\mured \propto \Phi^{-1}$.
Ultimately, at $\Phi = 0$ one would reach the ``bare'' limiting
value $\mured = l_B Z / R = 26$. At the higher volume fractions
$\Phi \geq 0.01$ charge renormalization yields $\Zeff < 30$,
giving rise to a reduced effective charge $\tildeZeff < 10$. In
this high--concentration regime, $\Zeff$ is fairly constant, and
approaching that of the $Z = 20$, $Z = 30$ systems. Hence, the
$\mured$ values are very similar, and again the mobility decreases
logarithmically with $\Phi$ (as do the experimental data).

\begin{figure}
\vskip 0.1in
\begin{center}
\includegraphics[clip, width=7.5 cm]{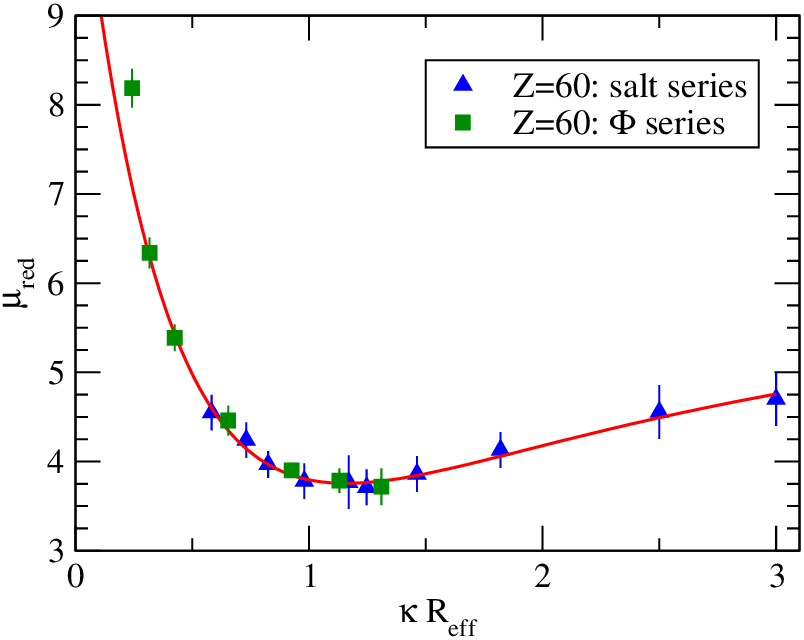}
\end{center}
\caption{Variation of the electrophoretic mobility of particles of
charge $Z=60$ in external field $Ee=0.1$ versus the reduced
screening parameter. The screening parameter was calculated from
the concentration of the free ions at different colloid volume
fractions and/or concentrations of added 1:1 salt. The curve is a
guide to the eye.} \label{fig:salt2}
\end{figure}

Although the behavior of $\mured$ is in general more complex, our
parametrization still remains valid. In Fig. \ref{fig:salt2} we
compare data obtained in a series of simulations at zero salt
concentration and increasing $\Phi$ to results of simulations at
constant $\Phi$ but increasing amounts of added salt. The data
coincide over a considerable region of $\kappa \Reff$. The salt series
curve continues with a slight increase, which is also a feature seen
in the classical electrokinetic works \cite{wiersema66a,obrien78a}.
Yet, we note that the minimum on the electrophoretic mobility can be
related to the general behavior of the ionic double layer for highly
charged colloids \cite{OrlandNetz}. The coincidence of the mobilities
obtained for matching $\tildeZeff$ and $\kappa \Reff$ again supports
our mapping postulate. Moreover, it clearly shows that the sort of
ions present in the cloud is of minor importance for the mobility. The
electrolyte effect, which leads to slowing down the particle drift, is
produced in one case by solely counterions and in another case by both
counterions and salt ions.  In both cases it is only the \emph{total}
ion concentration that matters.

Finally, we briefly comment on finite-size effects. The same
arguments that demonstrate hydrodynamic screening in systems with
salt \cite{Long,tanaka01a} apply here, too. Therefore finite-size
effects (as a result of image interactions) are expected to be
weak, i.~e. the single-colloid simulation should rather represent
the many-colloid situation at the same volume fraction. We have
explicitly tested this by increasing the number of particles at
fixed $\Phi$, which corresponds to a gradual transition from the
symmetry of a cubic crystal to an isotropic liquid (corresponding
to a bulk simulation \cite{lobaskin99}). The simulation results
(see inset in Fig.~\ref{fig:phi}) show that, within our numerical
resolution, the mobility is not affected by the positions of the
nearest neighbors. This is expected for a reasonably well--defined
double layer, and corroborates the general effective
single--particle picture, according to which the mobility is
governed by the shear stresses within that layer. Quite
analogously, one finds in the experiments that $\mu$ is remarkably
smooth at the freezing transition (Fig.~\ref{fig:salt1}).
Electrophoretic data on other particles as well as conductivity
data show similar behavior \cite{tom2,tom7}. Moreover, our
findings agree with recent numerical results for colloidal
dispersions, employing the numerical solution of the Stokes and
Poisson-Boltzmann equations \cite{Yamamoto,Chiang}.

In summary, we have studied the electrophoretic mobility of
colloidal particles in the counterion--dominated regime, which
sets in at finite particle volume fractions and low electrolyte
strengths. Based on the idea of static charge renormalization, we
suggest a set of effective control parameters, the reduced
effective particle charge $\tildeZeff$ and reduced screening
constant $\kappa \Reff$, controlling the static and dynamic
properties of the system. Our model is supported by a successful
matching of the LB/MD computer simulations of the primitive model
electrolyte and Laser Doppler velocimetry measurements with
similar $\tildeZeff$ and $\kappa \Reff$, whose non-rescaled system
parameters differ from each other by more than an order of
magnitude. We believe that these results and observations
constitute a first step towards understanding the physics of
colloid electrophoresis in the low-salt limit.

\acknowledgments{We thank J. Horbach, O. Vinogradova, and F.
  Carrique for stimulating discussions. This work was funded by
  the SFB TR 6 of the DFG.}


\end{document}